\documentclass[preprint,aps]{revtex4}

\usepackage{graphicx}
\usepackage{dcolumn}
\usepackage{bm}

\begin{document}
\title{Pair Wave Functions in Atomic Fermi Condensates}
\author{A. V. Avdeenkov \cite{byline} and J. L. Bohn}
\affiliation{JILA and Department of Physics, University of
Colorado, Boulder, CO 80309-0440}
\date{\today}

\begin{abstract}
Recent experiments have observed  
condensation behavior in a strongly interacting system of
fermionic atoms.  We interpret these observations in terms of 
a mean-field version of resonance superfluidity theory.
We find that the objects condensed are not bosonic molecules
composed of bound fermion pairs, but are rather spatially correlated
Cooper pairs whose coherence length is comparable to the mean
spacing between atoms.  We  propose experiments that
will help to further probe these novel pairs.
\end{abstract}
\maketitle

 Fermi condensates have been recently observed in dilute atomic gases,
first in $^{40}K$~\cite{Jin} and subsequently  in $^{6}Li$~\cite{Ketterle,Grimm}.  This new state of ultracold matter
represents a Fermi gas so strongly interacting that Cooper pairs
become correlated in physical space as well as in momentum space,
similar to the pairs  in high-$T_c$ superconductors.
Such materials are believed to exist in a ``crossover'' regime,
intermediate between  weak- coupling~(BCS) superconductivity
and Bose- Einstein condensation~(BEC) of tightly bound fermion pairs~\cite{Sademelo}.  An ultracold atomic Fermi system is an ideal
environment to explore the crossover regime, since the effective
interactions can be tuned via a magnetic field Feshbach resonance.
This possibility has led to various predictions based on a
'`resonance superfluidity'' theory of the gas~\cite{Timmermans,Holland,Ohashi}.  The BEC limit of the crossover regime was already achieved experimentally in the fall of 2003, with the creation of BEC of diatomic molecules composed of fermionic atoms 
~\cite{Greiner, Jochim, Zwierlein}.

Because of its close link to high $T_c$ superconductivity, the crossover
regime has been a topic of intense theoretical
investigation, beginning from its prediction 
~\cite{Sademelo,Eagles,Leggett}
and continuing through its recent adaptation to ultracold atomic 
gases~\cite{Timmermans,Holland,Ohashi,Falco,Xiong,Bruun,Ho}.  
A primary outcome of crossover theory is that the Cooper
pairs begin to become localized in space due to many-body
correlations as the interparticle interaction becomes large and 
attractive.  In the high-$T_c$ superconductor literature, these pairs
are referred to as ``pre-formed bosons, which can exist both
above and below their transition temperature to a Bose-condensed state.
The pairs in the crossover region are smaller than traditionally delocalized
Cooper pairs, yet are not rigorously bound molecules. In this Letter
we explore the link between pairs in the crossover regime and molecules by
explicitly constructing their wave functions for the conditions
of the experiment in Ref.~\cite{Jin}.  We find that the pairs
evolve smoothly into real molecules as the scattering length is
tuned from negative to positive values.  We also suggest
experiments whereby the spatial correlations of the pairs can be probed.
Note that a recent preprint comes to a similar conclusion for a
uniform (i.e., untrapped) Fermi gas \cite{Diener}.

This finding runs counter to the expectations of Refs.
\cite{Falco, Ketterle,Xiong}, where the pairs are identified with
actual molecules that are associated with the closed channel 
wave function in two-body scattering theory.  If this were 
the case, then the pair wave function would decay exponentially 
as a function of interparticle separation, regardless of
which side of the resonance it is on.  That this is
not the case will be demonstrated below.  Additionally,
Ref. \cite{Falco} identified the onset of the crossover regime
by setting the binding energy of the molecule, $\hbar^2/ma^2$
equal to the Fermi energy $\hbar^2(3\pi^2n)^{2/3}/2m$, where
$m$ is the atomic mass and $n$ is the number density of atoms.
Doing so, one finds that in this regime the scattering length
(hence the molecular size) is comparable to the interatomic
spacing, and the pairs are not yet recognizable as distinct
molecules.  They should rather be considered as spatially correlated
objects.

Accordingly, we study in this Letter the correlation length of
atom pairs.  Our starting point is 
 the resonance superfluidity approach~\cite{Holland1} adapted
within the Thomas-Fermi description~\cite{Me}. For concreteness,
we consider the  two- component Fermi gas of
$^{40}K$  near a Feshbach resonance between the  $|9/2-9/2>$ and
$|9/2-7/2>$ states~\cite{Regal}.  This system possesses a
Feshbach resonance whose zero-energy scattering is described
by an s-wave scattering length parametrized
by $a(B)=a_{bg}(1-w/\Delta B)$, with $a_{bg}=174 a_0$, $w=7.8$ G,
and $\Delta B$ is the magnetic field detuning in Gauss. For our numerical simulation we
have chosen the radial frequency $\nu_{r}=400Hz$ and the trap
aspect ratio  $\nu_{r}/\nu_{z}=80$, as in Ref. \cite{Jin}.

The primary objects of the resonance superfluidity theory
are the normal and anomalous densities, $\rho$ and $\kappa$,
representing the density of atoms in each species and of correlated 
pairs, respectively. $\kappa$ can be regarded
as the wave function of pairs.  Because we work in the local density approximation,
it is convenient to define these quantities as functions of the 
location ${\bf R}$ of each pair's center-of-mass in the
trap, and the relative momentum ${\bf k}$ of the pair.
  The equations of motion
for these quantities take the BCS-type form~\cite{Holland, Ohashi,Me}:
\begin{eqnarray}
\label{hfb} \nonumber \rho({\bf k,R})=n({\bf k,R})u^{2}({\bf
k,R})+ (1-n({\bf k,R}))v^{2}({\bf k,R}),
\\
\nonumber \kappa({\bf k,R})=u({\bf k,R})v({\bf k,R})(1-2n({\bf
k,R}))
\\
\nonumber E({\bf k,R})=\sqrt{h({\bf k,R})^2+\Delta({\bf R})^2};
\hspace{1.5cm}
 {u^{2}({\bf k,R}) \choose v^{2}({\bf
k,R})}=\frac{1}{2}(1 \pm \frac{h({\bf k,R})}{E({\bf k,R})})
\\
h({\bf k,R})=\frac{\hbar^2 k^2}{2m} + V_{mf}({\bf
R})+V_{trap}({\bf R}) - \lambda_c
\\
\nonumber
 V_{mf}({\bf R})=V_{bg}\rho({\bf R});\hspace{1cm}
\rho({\bf R})=\int \frac{d^{3}k}{(2\pi)^3}\rho({ \bf k,R})
\\
\nonumber \Delta({\bf R})=-V_{bg}\int
\frac{d^{3}k}{(2\pi)^3}\kappa({\bf k,R})-g\phi({\bf R});
\hspace{1.5cm}
 \phi({\bf R})=\frac{g}{2\lambda_c
-\nu} \int \frac{d^{3}k}{(2\pi)^3}\kappa({\bf k,R})
\end{eqnarray}
where 
 $n({\bf k,R})=(exp(E({\bf k,R})/k_{B}T)+1)^{-1}$ is the Fermi-Dirac distribution,
 $V_{mf}({\bf R})$ is the mean field potential, $\Delta({\bf R})$ is the energy gap,
$V_{bg}=4 \pi \hbar^2 a_{bg}/m$, $a_{bg}$ is the
background~(non-resonant) contact interaction, $\nu=(B-B_{0})\Delta
\mu $ is the detuning in energy units, $g=\sqrt{V_{bg} \Delta B
\Delta \mu}$ is a coupling representing the conversion of free fermions
into pairs, $\Delta B$ is the field width of the resonance,
$\Delta \mu $ is the magnetic moment difference between two hyper-
fine levels of the  two-component Fermi gas, $\lambda_c$ is the
chemical potential, and $V_{trap}$ is the
external atomic  harmonic trapping
potential. The chemical potential is fixed by conservation
of the mean number of atoms $N$.
In this theory, a ``molecular field'' $\phi$ is introduced to 
simplify the theoretical
description of free fermions transforming into spatially correlated pairs.  Notice that $\phi$ is not a distinct physical entity,
but is determined once $\kappa$ is known.

 Though derived for interacting fermions, equations~[\ref{hfb}] 
can also be applied in the BEC
limit~\cite{Sademelo,Babaev,Pistolesi} In this limit it is well- known that ~($ v^{2}({\bf k,R}) << 1 $) and that the BCS equation for the gap reduces to the Schr\"{o}dinger equation for the relative 
motion of two interacting bosons, with energy eigenvalue
$2\lambda_c=\hbar^2 /ma^{2}$ representing the binding energy
of the (rigorously bound) bosonic molecules~\cite{Pistolesi,Viverit}.  After a simple derivation we
can extend this result to the case of trapped atoms.  Then the
pair distribution function becomes, in the BEC limit,
\begin{eqnarray}
\label{kappak} \kappa({\bf k,R})=\frac{m\Delta({\bf R})a^{2} }{k^2 a^2
+1}= \frac{\sqrt{8\pi a^3}}{k^2 a^2 +1}*\Phi({\bf R}),
\end{eqnarray}
where $\Phi({\bf R})=\sqrt{\frac{m^2 a}{8\pi}}\Delta({\bf R}) $ as
 derived in~\cite{Pieri}. This reference also demonstrated that HFB
equations transform into the Gross- Pitaevskii equation
 and that $\Phi(\bf R)$ serves as a solution for a
molecular BEC interacting through a repulsive interaction with
scattering length $a_B =2a$. This result holds in the BEC limit,
where  $\Delta / |
\lambda_{c}| <<1$.  (A more careful analysis, following~\cite{Petrov},
 will give corrections, but this is not the main goal
of this Letter.)
 Moreover, the first term on the right-hand side of Eqn.~(\ref{kappak})
 is the Fourier
transform of the molecular wave function $\frac{1}{\sqrt{2 \pi a}}
\frac{e^{-r/a}}{r}$ in  the relative coordinate $r$. In this way the same 
wave function $\kappa$ that represents Cooper pairs on the 
BCS side of the resonance actually represents a condensate of real molecular bosons on the BEC side.  In general
this molecular wave function depends on $\bf R$, 
meaning that the gas may contain molecules in its high-density
center, but correlated Fermi pairs at its lower-density
periphery. For example in this case  twice the chemical
potential is not quite the molecular binding energy but slightly
depends on density~\cite{Viverit}. 

The number of pairs can be calculated from the anomalous density as $ N_{b}=\int
d{\bf R} d{\bf k} \kappa({\bf k,R})^{2} $. In the BEC limit
$\kappa({\bf k,R})^{2} \Rightarrow \rho({\bf k,R})$ for small temperatures,
 which means
that  almost all atoms are paired. Moreover, using (\ref{hfb},
\ref{kappak}) it is easy to check that in this limit the density
of pairs  transform into the density of real
molecules:
\begin{eqnarray}
\rho_{b}({\bf R}) = \int d{\bf k} \kappa({\bf k,R})^{2}= \Phi({\bf
R})^2
\end{eqnarray}
Thus the same function  $\kappa({\bf k,R})$ describes the density of
Cooper pairs away from resonance, pairs in the crossover regime,
and molecules on the BEC side of the resonance as the detuning is varied.

The many- body physics of the
crossover regime can be quantified in terms of a 'smooth'
parameter such as the pair coherence length, usually defined as the rms radius of the pair~\cite{Pistolesi}:
\begin{eqnarray}
\label{length} \xi^{2}({ \bf R})=\int d{\bf r}
\kappa({\bf r,R})^{2} r^{2} / \int d{\bf r} \kappa({\bf r,R})^{2}
\approx (k_{F}({\bf R})/ m \pi \Delta({\bf R}))^2
\end{eqnarray}
Using the above result, it is clear that in the BEC limit $<r^{2}>=a^{2}/2$ in the center of the trap (note that the
size of a molecule is usually taken instead as the mean value of $r$,
$<r>=a/2$). On the BCS side of the
resonance $\sqrt{\xi^{2}({ \bf R})}$ defines the 'size' of the
Cooper pair. Thus the calculation of the coherence length  gives us
an insight of how the pairs evolve in the crossover regime.
 
\begin{figure}
\centerline{\includegraphics[width=0.9\linewidth,height=1.08\linewidth,
angle=-90]{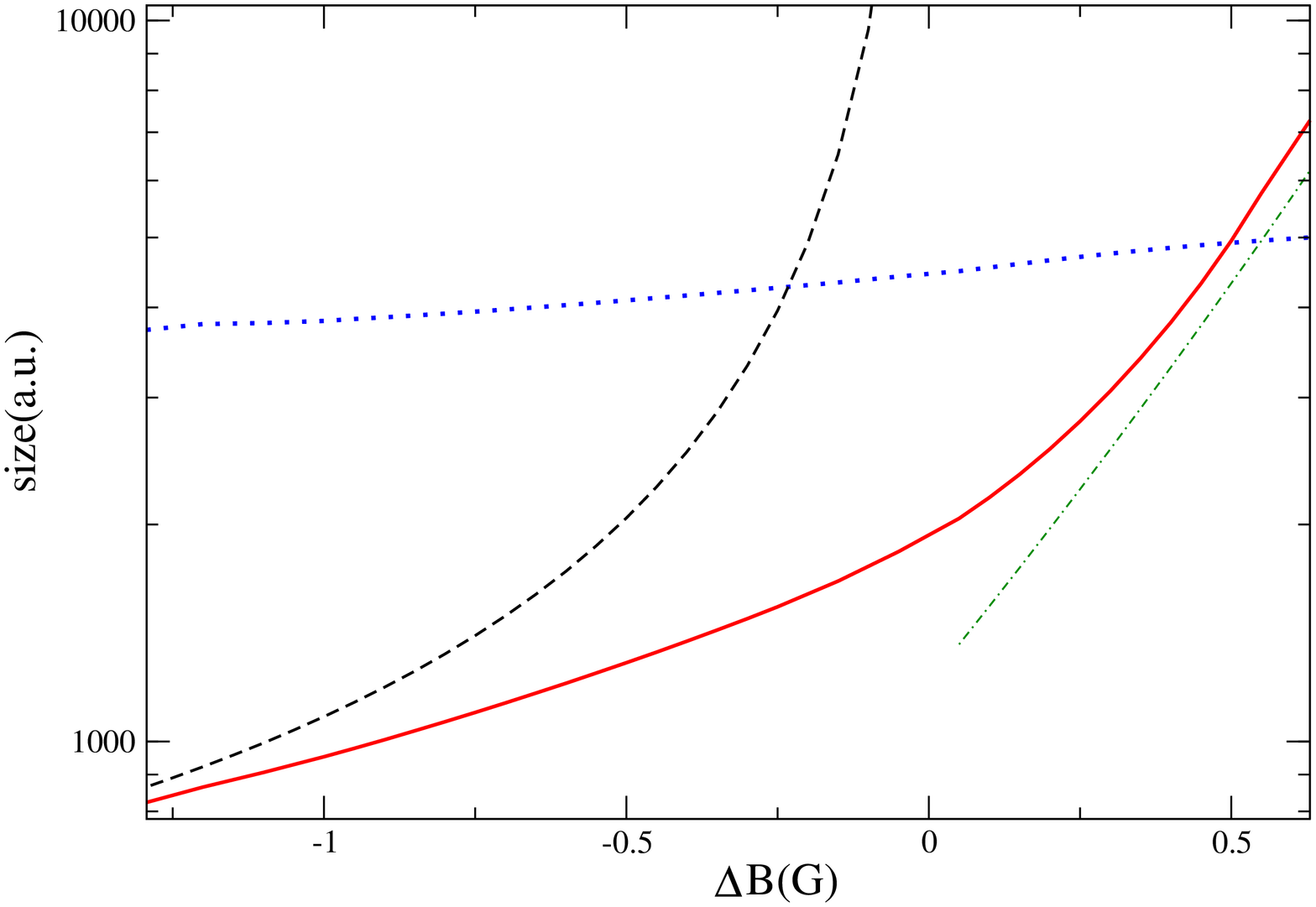}}
\caption{Coherence length versus magnetic field detuning (solid line)
for $^{40}$K atoms in the JILA experiment, in the center of the trap.  
For comparison, the dashed
curve represents the rms molecule size $a/ \sqrt{2}$ corresponding to atoms with a scattering length  $a=a_{bg} -
\frac{g^2}{\nu}$. The dash-dotted curve is the BCS limit of the coherence length.
The  dotted curve represents the interparticle
distance in the center of the trap.
 The trap aspect ratio $1/ \lambda =80$, the temperature is $T=0.08T_{F}$,
 and the trap contains $N=5 \times 10^{5}$ atoms.
} \label{size}
\end{figure}

We present the coherence length versus detuning in Fig.\
\ref{size} (solid line) for the trap aspect ratio, number of atoms,
and  temperature of
the JILA experiment~\cite{Jin}. For detunings $\Delta B > 0.5$ G the
coherence length approaches the familiar BCS result (dash-dot line).
For negative detunings $\Delta B < -1$ G on the BEC side of the
resonance, the molecular size approaches the size evaluated
from two-body theory (dashed line).  In between, the coherence length
varies smoothly, illustrating the gradual evolution of pairs
into molecules.  The sizes of these objects remain finite
across the resonance, in spite of the divergent behavior of
the scattering length.  This size suppression is the result of
many-body physics in the unitarity limit of $k_Fa>1$, where
the physics of the gas is expected to saturate and to depend only
weakly on the scattering length.

\begin{figure}
\centerline{\includegraphics[width=0.4\linewidth,height=0.5\linewidth,angle=-90]{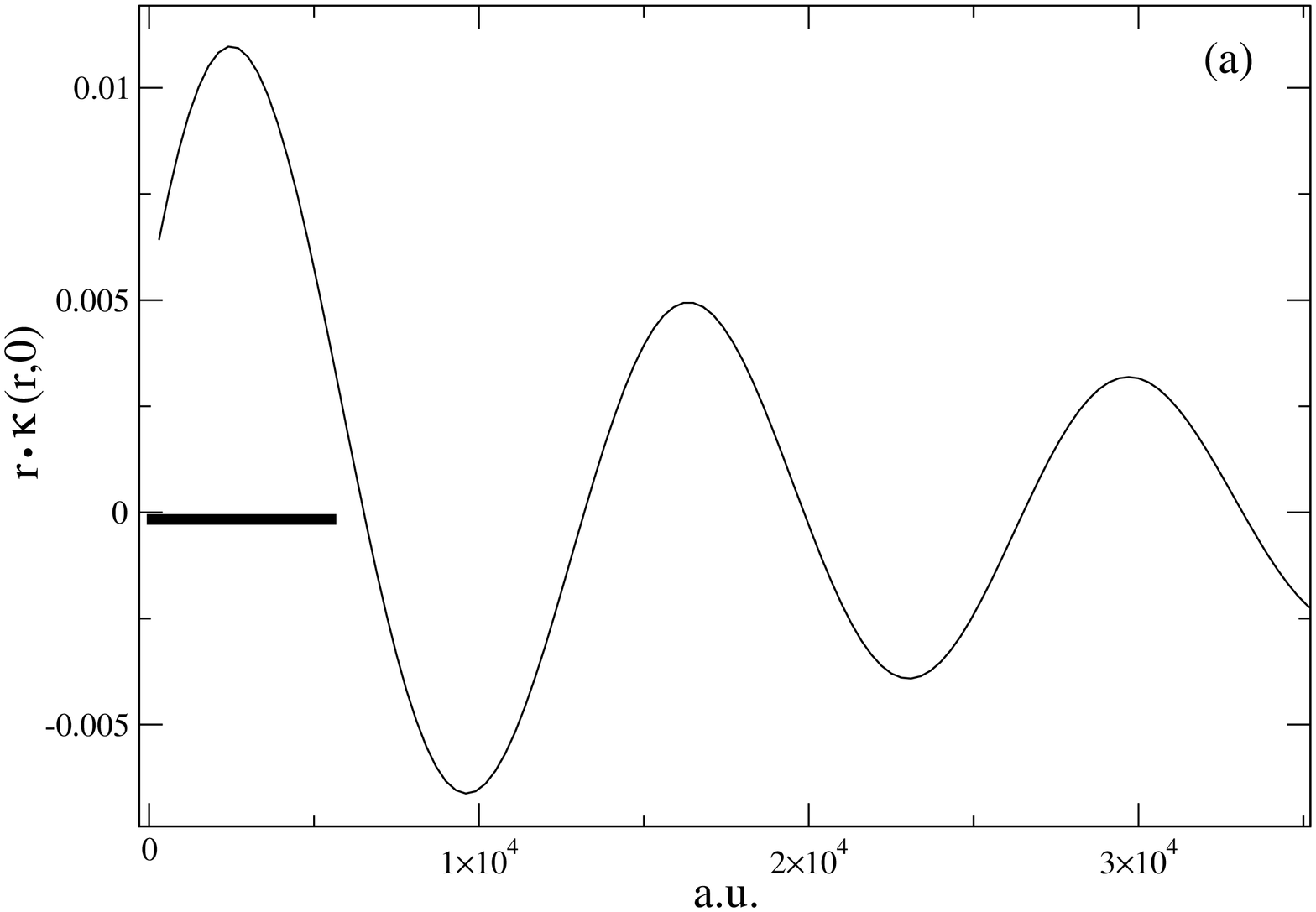}}
\centerline{\includegraphics[width=0.4\linewidth,height=0.5\linewidth,angle=-90]{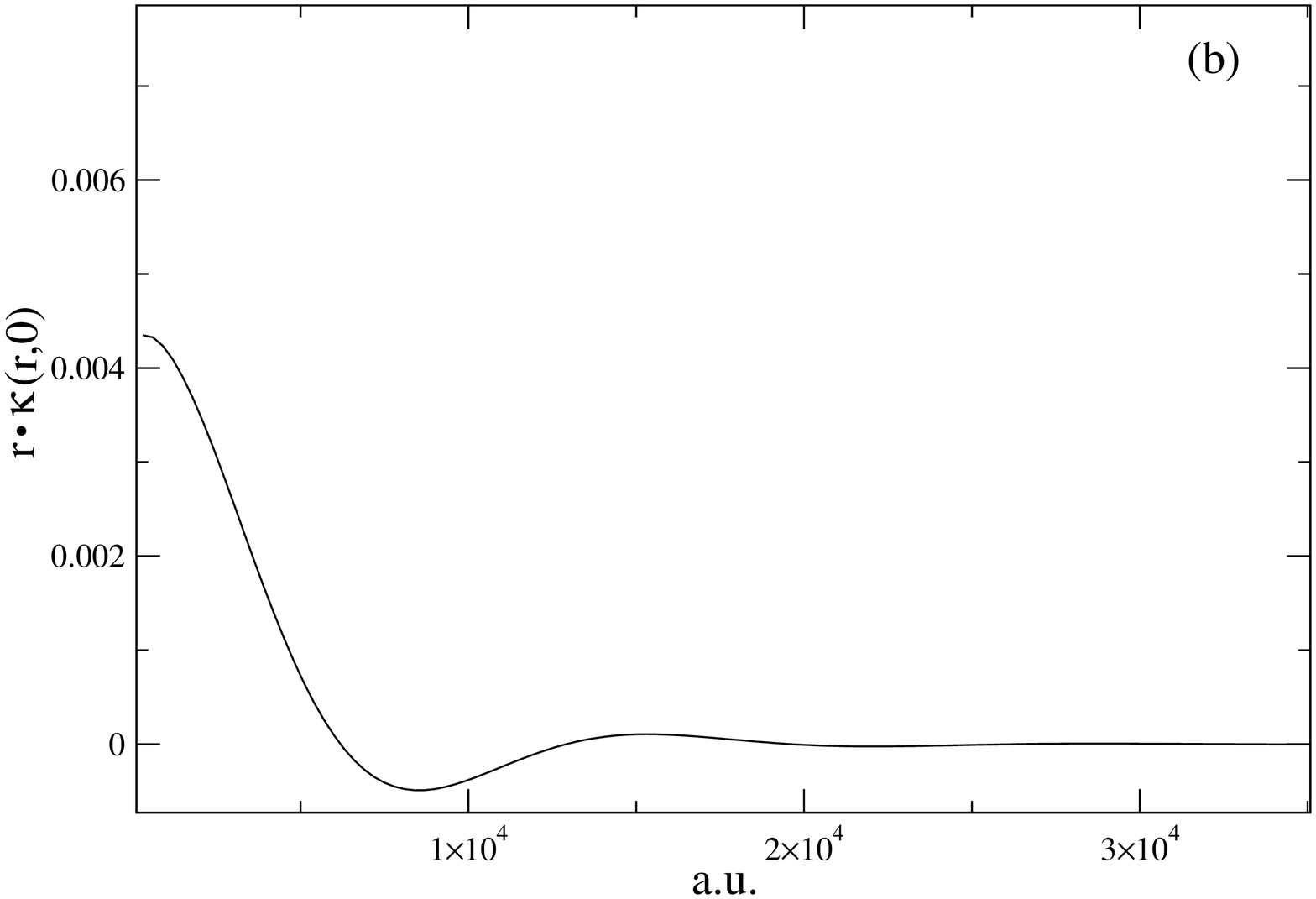}}
\centerline{\includegraphics[width=0.4\linewidth,height=0.5\linewidth,angle=-90]
{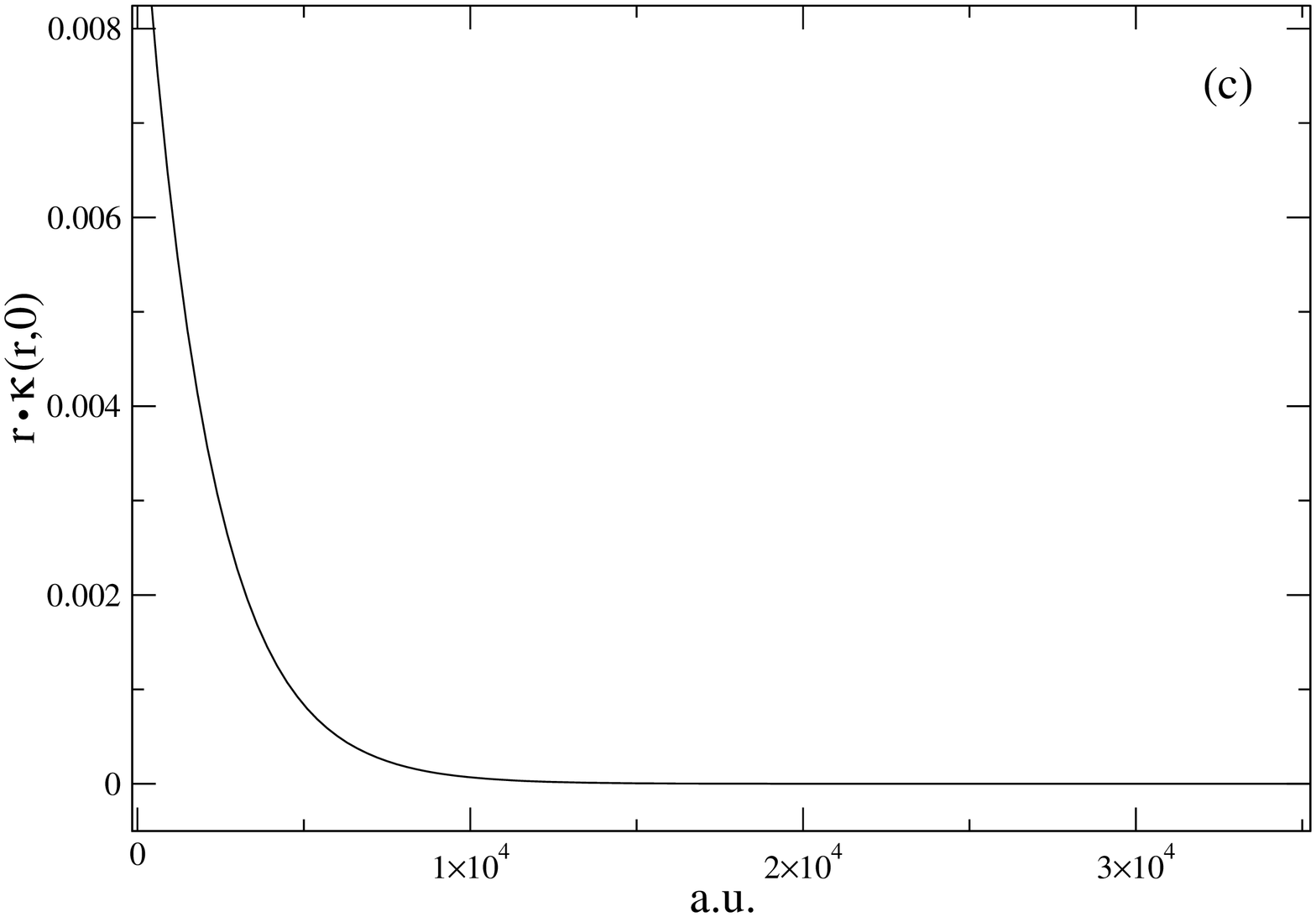}}
\caption{Two-body correlation function $r \cdot \kappa({\bf r},{\bf R}=0)$
versus interparticle separation $r$ in the center of the trap considered 
in Fig. 1.  The panels correspond to the detunings $\Delta B=1.0$G (a),
$\Delta B=0.1$G (b), and $\Delta B=-0.5$ G (c).  On the negative detuning
side of the resonance, the pairs are true molecules.
For comparison, the solid bar in (a) shows the interparticle distance 
in the center of the trap. } \label{wf}
\end{figure}

To illustrate in greater detail the smoothness of the transition
between pairs in the crossover regime and molecules, we consider their
wave functions, as shown in~Fig.\ \ref{wf}.  This figure shows
pair wave functions $r \cdot \kappa({\bf r},{\bf R}=0)$ in the center
of the trap, for detunings corresponding to ``ordinary'' Cooper pairs 
($\Delta B=1.0$G, a), pairs in the crossover regime
$\Delta B=0.1$ G, b), and molecules $\Delta B= -0.5$G, c).
Wave functions of the pairs decay away on a length
scale set by the coherence length, but in an oscillatory way
reminiscent of a damped harmonic oscillator.  This behavior
is a many-body effect, and in fact the scale of the 
oscillation is set by the interparticle distance (solid bar).
The relative motion of true molecular bound states of course
decay strictly exponentially, as in Fig.2c).
For small detuning, however, the correlation length becomes comparable
to the molecular size, and the ringing wave functions begin
to resemble overdamped oscillators, i.e., they  
decay exponentially (Fig.2b).  
In this way the character of the pair wave functions 
evolve smoothly into molecular wave functions.
 This interpretation is somewhat complicated by the fact that the shape of pair wave function
strongly depends on the trap geometry. Even for small detunings
but far from the trap's center the coherence length is still large.

\begin{figure}
\centerline{\includegraphics[width=0.9\linewidth,height=1.08\linewidth,angle=-90]{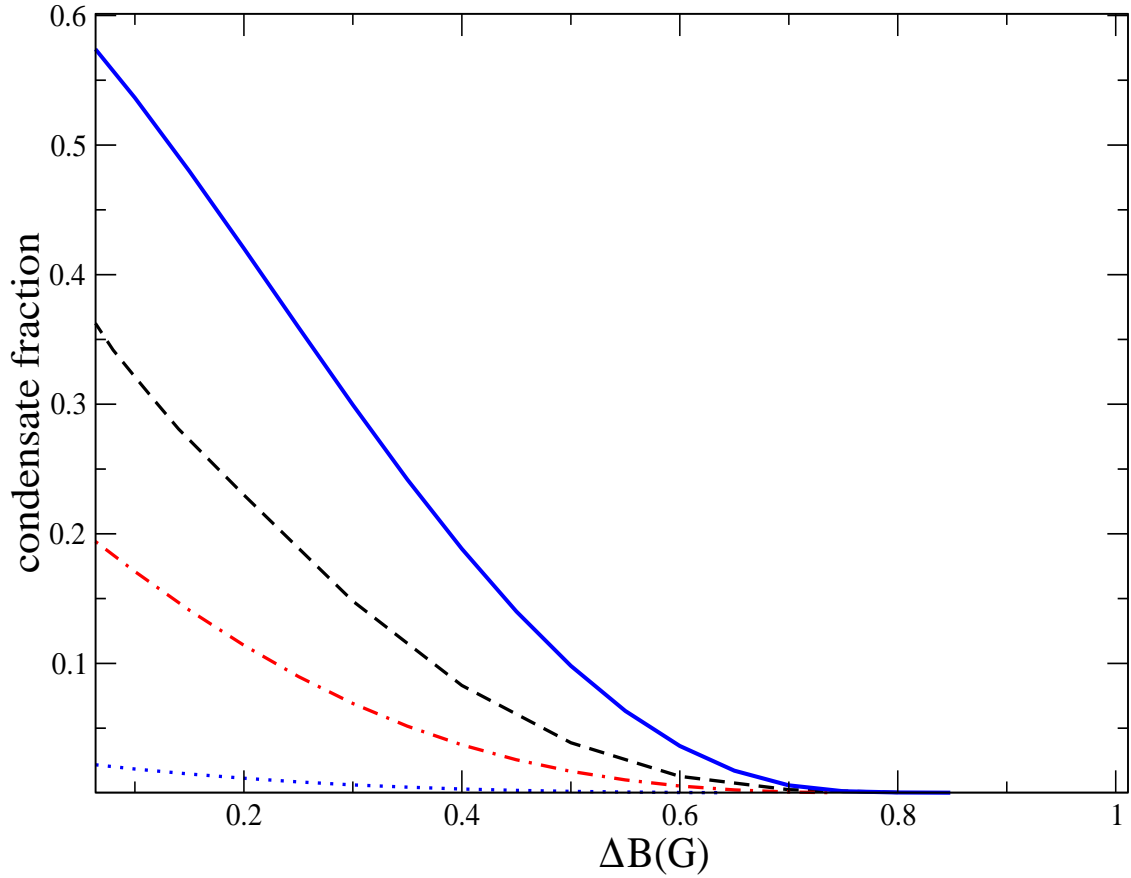}}
\caption{Condensate fraction versus detuning, as measured by
projecting onto molecules, according to Eqn. (5), for the trap
considered in Fig. 1.  The molecules
projected onto correspond to atomic interactions with scattering lengths 2000, 1000,  and 200 $a_0$
(dashed, dot-dashed and dotted lines respectively).
The solid line is the {\it in situ} condensate fraction,
given by  $N_{b}/(N/2)$.
} \label{frac}
\end{figure}

 In order to qualitatively  understand the JILA 
experiment~\cite{Jin} we now consider the positive detuning, BCS 
side of Fig.\ \ref{size}. We  see that at a detuning of around $0.5G$ the 
size of the pairs becomes comparable to the interparticle 
distance (dotted line in Fig.\ \ref{size}).  This criterion marks
the crossover regime, where the atom pairs are not momentum-correlated
objects like BCS Cooper pairs, nor are they yet full-fledged molecules.
Significantly, this detuning is approximately where a condensate
fraction can be observed in the JILA experiment \cite{Jin},
implying that the condensed objects consist of correlated pairs
rather then real molecules.
 To estimate the condensate fraction, we assume
for simplicity that all pairs
are Bose condensed at experimental temperatures, so that
the condensate fraction is simply $N_{b}/(N/2)$. The true condensate
fraction presumably depends on the (unknown) interaction between
the [airs.  This
{\it in situ} condensate fraction is presented as a function
of magnetic field detuning in Fig.\ (\ref{frac}) (solid line).
This fraction becomes significant only for detunings less than
about 0.5 Gauss from resonance,
just where the size of the pairs becomes comparable
to the interparticle spacing (compare Fig.1).  
   The condensate fraction is quite large
near zero detuning. In the ideal case of a uniform gas this fraction
would be $1$ on resonance, but it is generally smaller for a trapped gas.

In the JILA experiment, the Fermi condensate is not directly
imaged, but rather is probed by 
a magnetic field sweep that converts the atoms into
molecules.  This sweep is fast enough that it does not affect
the many-body properties of the gas, but slow enough that
atoms are efficiently gathered into molecules.  The final detuning
is far below resonance ($\sim -10$G in Ref.\cite{Jin}), so
that the molecules are far smaller than the pairs
that are being probed.  An infinitely fast sweep that 
instantaneously projects pairs onto molecules would therefore
not yield a significant number of molecules.  The condensate 
fraction, however, could still be a significant fraction of
unity \cite{Diener}.

In the present calculation, we do not treat the time dependence
of the magnetic field, and therefore cannot model the experiment
as performed.  We can, however, suggest another experiment that
could probe the crossover regime more fully.
Let us consider a hypothetical experiment where it would be possible to apply an
infinitely  fast sweep from  a positive detuning $\Delta B_{BCS}$ to a final 
detuning $\Delta B_{BEC}$, i.e., literally projecting pairs onto
molecules.
The final condensation fraction $f$
observed by expansion and imaging will
then be defined as a product of two probabilities: the first is
 the projection of the  pair wave function 
$\kappa({\bf r, R})$ onto the molecular wave
function~(\ref{kappak}), normalized by $N_b$; the second
is the fraction of atoms that are paired, ($N_{b}/(N/2)$):
\begin{eqnarray}
\label{overlap} f(\Delta B_{BEC},\Delta B_{BCS})=\left( \int d{\bf r} d{\bf
R} \kappa({\bf r, R}) \frac{1}{\sqrt{2 \pi a}}
\frac{e^{-r/a}}{r}\Phi({\bf R}) \right)^{2} * \frac{N_{b}}{N/2}.
\end{eqnarray}
 This projection depends not only on the mapping of the
  wave function of a pair onto the wave
function of the molecule but also  on the condensate wave function as
$\Phi({\bf R})$.  The fermionic condensate wave function
$\kappa({\bf r, R})$ cannot be so easily separated as the product
of center-of-mass and relative functions as in the BEC case~(\ref{kappak}).  It should be said
that even in BEC case the molecular wave functions
in~(\ref{kappak}) will depend on $\bf R$, which means that the
molecular size will be different from point to point in the trap.
But this dependence is quite weak especially for large negative
detunings, and is therefore neglected. In the case of a
large positive detuning $\Delta B_{BCS}$ the  size of a  Cooper pair is
considerably larger than the size of a molecule and the number of
pairs $N_b$ is quite small itself so the overlap
integral~(\ref{overlap}) will be quite small. It is  clear that
the observed condensate fraction will depend on the geometry of the trap as
well as on the detuning $\Delta B_{BEC}$ of the final point of the
sweeping.

 We have calculated the condensate fraction, as seen by this
projection technique,  for
 three different target molecules defined by $\Delta B_{BEC}$, corresponding
to molecules of sizes 
$a/2=1000a_{0}, 500a_{0}$ and $100a_{0}$ ~( Fig.\ \ref{frac}).
As anticipated, the condensate fraction measured in this way
would be smaller if the pairs are projected onto smaller molecules.
 Thus there are two conditions required to support
a large observed condensate fraction: first the  $\Delta B_{BCS}$ detuning on
the BCS side of the resonance should be small enough to support
a considerable number of pairs compared to the total
number of  atoms; and second the $\Delta B_{BEC}$ on the BEC side of the
resonance should be chosen so that the corresponding scattering length
will not be very different from the coherence length corresponding
$\Delta B_{BCS}$ . The second condition means that the size of the pair
should be comparable with the size of the molecule. For the
experiment with $^{40}K$ atoms these conditions are fulfilled  for
$\Delta B_{BCS}<0.6G$ and  $\Delta B_{BEC}>-1G$. Of course these results
strongly depend on the geometry of the trap and the temperature.
An experimental
map of $f(\Delta B_{BEC},\Delta B_{BCS})$ should prove quite illuminating
as a probe of the length scales and condensation fractions
in the crossover regime.

In conclusion, we found that the recent experiment~\cite{Jin}
can be explained semi-quantitatively by counting the number of Cooper
pairs  on the BCS side of the Feshbach resonance. 
We suggested a new scenario to probe the crossover regime 
 by mapping the  condensate  of fermionic
pairs on the BCS side of the resonance onto molecules on the BEC side. 
We found that a considerable condensate fraction can be observed when 
the coherence length of the pairs is on the order 
of the interparticle distance.

This work was  supported by the NSF.  We acknowledge useful discussions with
C. Regal, M. Greiner, and D. S. Jin.

\end{document}